\documentclass[11pt]{article}

\usepackage[a4paper,margin=1in]{geometry}
\usepackage{amsmath,amssymb,amsthm,mathtools}
\usepackage{bm}
\usepackage{graphicx}
\usepackage{hyperref}
\usepackage{enumitem}
\usepackage{microtype}

\hypersetup{
  colorlinks=true,
  linkcolor=blue,
  citecolor=blue,
  urlcolor=blue
}

\theoremstyle{definition}

\theoremstyle{remark}

\title{Will a Large Complex System be Stable? Revisited}

\author{
Michael A.S. Thorne\\
  British Antarctic Survey, Cambridge, UK\\
  \texttt{mior@bas.ac.uk}
}
\date{}

\begin{document}
\maketitle
\begin{abstract}
Over fifty years ago, Robert May applied random matrix theory to show that as ecological systems grow in size, stability decreases. What emerged from this and the critique that followed was decades of what has been called the complexity-stability debate. However, decades of critique over the assumptions that Robert May applied in carrying out his analysis have not been enough to fully dispel the strength of his conclusion and close the debate. Drawing on a mathematical approach that had not yet been fully developed in the early 70s, it is possible to revisit the argument without the use of random matrix techniques, and provide more detailed understanding of the mechanisms that play a deciding role in stability of ecological systems, countering the broad conclusion that led to the complexity-stability debate.
\end{abstract}

In his classic 1955 paper on food-webs \cite{MacArthur1955}, Robert MacArthur made the observation that systems that have more alternative pathways for energy transfer are more stable, reinforcing longstanding assumptions about the increased stability of more complex systems \cite{Elton1927,Odum1953}. Based on computer simulations in 1970 by Gardner and Ashby that showed how stability was affected by changes in the connectance of a system \cite{GardnerAshby1970}, Robert May applied random matrix theory to show more generally that as random matrices increase in size, stability decreases \cite{May1972,May1973}. May considered a linearized community matrix, or Jacobian, $J$, of an ecological system, $J = -dI + A$ with uniform self-regulation $d > 0$, and where the interaction strengths $a_{i \ne j}$ of the random matrix $A$ had the properties $\mathbb{E}[a_{ij}] = 0$, variance $\sigma^2$, and connectance between the species $C$. Using random matrix theory, he showed that the eigenvalues of $A$ lie approximately in a disk of radius $\sigma \sqrt{nC}$, where $n$ is the matrix size, or number of species. For stability this would require that $d > \sigma \sqrt{nC}$, which as the size, and therefore complexity, of the system grows, would require increasingly larger self regulation, leading to the conclusion that more complex systems are unstable. 

After decades of critique levelled at May's approach  \cite{CohenStephens1978,Pimm1979,Pimm1982,CohenNewman1985,DeRuiter1995,McCann1998,Neutel2002,Neutel2007}, arguing that introducing more realistic ecological structure into the governing equations and therefore the Jacobians is in fact stabilising, Allesina and Tang \cite{AllesinaTang2012}, forty years after May's paper, applied a very similar random matrix approach and were able to show that predator-prey structure does effect the distribution of the eigenvalues. By introducing a correlation function 
$\rho = corr(a_{ij},a_{ji}) < 0$ 
where $a_{ij}$ and $a_{ji}$ have different signs due to the predator-prey structure ($a_{ij} a_{ji}\le0$), they could reframe May's result as $d > \sigma\sqrt{nC}(1+\rho)$, which deforms the eigenvalue distribution from a circle to an ellipse, shifting the maximal real part of the spectrum leftward, which acts to increase stability.

In 1892, Aleksandr Lyapunov published his foundational work on the stability of dynamical systems \cite{Lyapunov1892}, introducing the direct method for determining the stability at equilibrium without explicitly solving the system, through what are now called Lyapunov functions.\\
\indent Subsequently developed throughout the twentieth century \cite{LaSalle1960,Kalman1963,Karmarkar1984,NesterovNemirovskii1994}, a major computational breakthrough came in 1994 with the work of Boyd \textit{et al.} \cite{Boyd1994}, who showed that the search for quadratic Lyapunov functions can be formulated as a linear matrix inequality (LMI) and solved via convex optimisation. This transformed Lyapunov’s existence theorems into practical, algorithmic stability criteria (or certificates in control-language terminology) for large-scale interconnected systems.

Reframing May’s result in the language of an LMI enables one to replace spectral radius bounds with constructive Lyapunov certificates. This extends the analysis beyond random ensembles to structured and deterministic interaction matrices. Given the linear system $\dot{y}=Jy$, then the Jacobian $J$ is Hurwitz (i.e. all eigenvalues have negative real part) if and only if there exists a symmetric positive definite matrix $P\succ 0$ such that $J^\top P + PJ \prec 0$. This follows from the quadratic Lyapunov function $V(y)=y^\top P y$, whose derivative along trajectories is $\dot{V}(y) = y^\top(J^\top P + PJ)y$. Asymptotic stability of the linear system is then equivalent to the feasibility of the linear matrix inequality $J^\top P + PJ \prec 0$.

In the Lyapunov condition $J^\top P + PJ \prec 0$, the matrix $P$ does not alter the ecological system itself and the Jacobian $J$ is fixed. Rather, $P$ defines the quadratic function $V(y)$ which determines the metric in which perturbations are measured. Different choices of $P$ therefore correspond to different geometric lenses through which stability can be assured (or certified). What changes across certificates is not the dynamics, but the structural explanation of why stability holds.

One may therefore construct different certificates to derive structural criteria. Returning to the Jacobian $J$, we can, with the LMI apparatus, work with non-uniform diagonal values, which many of the critiques of the random matrix approach have suggested has been missing. Therefore, we can rewrite $J = -D + A$, where $-D$ ($D=diag(d_1,\dots,d_n), d_i>0$) encodes non-uniform self-regulation. Similar to the analysis carried out by Allesina and Tang \cite{AllesinaTang2012}, we can define the predator-prey sign structure such that for every unordered pair $\{i,j\}$ with $i\neq j$, $A_{ij}A_{ji} \le 0$ (if $|A_{ij}| > 0$ then $|A_{ji}| > 0$). A convenient parametrization is $A_{ij} = b_{ij}\ge 0$ and $A_{ji} = -c_{ij}\le 0$.

Allowing a full positive definite $P \succ 0$ removes geometric restrictions, which is equivalent to the Hurwitz stability of $J$. However, constructing certificates with more conservative constraints can be structurally revealing. As a way of illustrating the structural flexibility of the Lyapunov framework, we can consider three certificates of increasing generality. The first, which may be termed the Euclidean certificate, fixes $P=I$ (the identity), yielding the condition $J^\top+J \prec 0$. In this case, stability is determined directly from the symmetric part of $J$, therefore it depends only on the symmetric imbalance relative to self-regulation. This is the most restrictive but also the most transparent geometric choice. A looser constraint is obtained by allowing 
$P$ to be diagonal, $P = diag(p_1,p_2,...,p_n), p_i>0$. The corresponding condition involves weighted pairwise terms $(J^\top P + PJ)_{ij} = p_i J_{ij} + p_j J_{ji}$, effectively replacing raw interaction imbalance with proportional, species-weighted imbalance. This certificate allows species-specific rescaling in the stability metric, capturing whether proportional balance across heterogeneous species can render the system stable. The third formulation, rather than simply certifying stability, quantifies the rate at which perturbations decay by bounding the full certificate: 
$J^\top P + PJ \preceq -2\alpha P$. The parameter $\alpha$ provides a lower bound on exponential convergence and therefore measures the strength or resilience of stability. Together, these certificates form a sort of hierarchy from direct symmetric dissipation, to proportional balance under species heterogeneity, to quantitative resilience under perturbation.

To explore in more detail the  Euclidean choice, we can decompose $A$ into its symmetric ($S$) and skew ($K$) components (i.e. $A = S + K$), which can reveal how each part contributes to stability:
\begin{equation*}
S := \frac{A + A^\top}{2},\quad
K := \frac{A - A^\top}{2}.\quad
\end{equation*}
Then $S=S^\top$ and $K^\top=-K$. Therefore, for each $i\neq j$, 
\begin{equation*}
S_{ij} = \frac{A_{ij}+A_{ji}}{2} = \frac{b_{ij}-c_{ij}}{2}, \quad
K_{ij} = \frac{A_{ij}-A_{ji}}{2} = \frac{b_{ij}+c_{ij}}{2}.
\end{equation*}
Therefore $S_{ij}$ is the pairwise imbalance (difference) between predator and prey effects and $K_{ij}$ is the pairwise interaction magnitude (sum) and can be large. We can then write the certificate by introducing these separate components (and setting $P=I$), 
\begin{align*}
\dot V(y)
&= y^\top(J+J^\top)y \nonumber\\
&= y^\top\big((-D+A)+(-D+A)^\top\big)y \nonumber\\
&= y^\top\big(-2D + (A+A^\top)\big)y \nonumber\\
&= 2\,y^\top(S-D)y.
\end{align*}
Because $K$ is skew-symmetric, $y^\top K y=0$ for all $y$, so the skew component does not contribute to $\dot V$ under this certificate. This is meaningful, since it shows explicitly that stability is governed by the symmetric residual imbalance while the skew-symmetric component represents reciprocal predator–prey exchange without directly affecting stability. Since $S-D$ is symmetric, $S - D \prec 0$  holds if and only if $\lambda_{\max}(S-D) < 0$. A sufficient condition is $\lambda_{\max}(S) < d_{\min} := \min_i d_i$. Therefore, in the Euclidean certificate, instability can arise only if symmetric interaction imbalance overwhelms the weakest self-regulation coefficient. 

In the diagonal certificate ($P = diag(p_1,p_2,...p_n), p_i>0$), this leads 
the original imbalance $b_{ij} - c_{ij}$
to be replaced by the weighted imbalance
$p_i b_{ij} - p_j c_{ij}$. Stability requires that this weighted imbalance, aggregated across all neighbours of species $i$, be dominated by the weighted self-regulation $p_i d_i$. Stability then depends not on absolute symmetry of interactions, but on whether directional asymmetries can be proportionally balanced relative to species scale. Thus, the diagonal certificate uniquely reveals whether heterogeneity in species influence or damping can absorb interaction imbalances that would appear destabilizing in the Euclidean case.

Finally, instead of merely requiring feasibility of the Lyapunov inequality one can strengthen the condition to $J^\top P + PJ \preceq -2\alpha P$, for some $\alpha > 0$. This strengthened inequality does not simply assert that trajectories decay; it quantifies the rate at which they do so. That is, if $V(y) = y^\top P y$, then $\dot V(y) = y^\top (J^\top P + PJ) y \le -2\alpha y^\top P y = -2\alpha V(y)$ which satisfies $V(t) \le e^{-2\alpha t} V(0)$, showing that perturbations decay at least exponentially at rate $\alpha$. The parameter $\alpha$ therefore represents a certified exponential decay rate for deviations from equilibrium. Ecologically, $\alpha$ measures how rapidly the system returns to equilibrium following a disturbance. A large $\alpha$ corresponds to fast recovery and strong damping of perturbations, while a small $\alpha$ indicates slow recovery and fragility. In this sense, $\alpha$ provides a quantitative measure of resilience rather than a mere binary indicator of stability. Two systems may both be locally stable in the sense that all eigenvalues of $J$ have negative real part, yet one may have eigenvalues clustered near zero and recover slowly, while another has eigenvalues far into the left half-plane and recovers rapidly. This formulation makes the distinction explicit. This perspective opens up structural ecological questions that do not arise in a purely spectral analysis. One can ask how predator-prey sign structure affects not only whether the system is stable, but how strongly stable it is. One can examine how $\alpha$ scales with increasing diversity. This shifts the focus from the existence of stability to the strength and robustness of stability, providing a refined lens through which to study the complexity-stability relationship.

Recasting May’s question of stability in the language of Lyapunov linear matrix inequalities does not replace random matrix analysis with an alternative asymptotic theory, although such asymptotic results could be developed through this approach. Rather, it reframes the complexity–stability relationship through the geometry of Lyapunov stability certificates. In the classical analyses of May and the refinement of Allesina and Tang, stability is characterised through the spectral properties of random interaction matrices and their behaviour in large ensembles. The LMI formulation instead asks whether there exists a metric in which perturbations decay, shifting the emphasis from typical spectral behaviour to the structural conditions that allow a particular system to remain stable. 
This perspective has two important consequences. First, it applies directly to deterministic and structured interaction matrices, allowing heterogeneous self-regulation and realistic ecological interaction patterns to be incorporated without altering the mathematical machinery. Second, the convex structure of the Lyapunov inequality provides a computationally tractable way to test stability and quantify resilience. In this setting, the classical random matrix results of May can be interpreted as describing how stability certificates typically fail in unstructured systems as complexity increases. Rather than negating the results of May, the LMI approach opens up the possibility of investigating how interaction structure, heterogeneity, and species-specific regulation combine to produce stability. Through the LMI machinery, the question is no longer, do larger systems become unstable, but rather what structural characteristics enable systems to remain stable as they grow.

\end{document}